\documentclass[showpacs,aps,prd,superscriptaddress,nofootinbib,floatfix,amsmath,amssymb]{revtex4}
\usepackage{graphicx}
\def\mhc{m_{H^\pm}}
\def\dhch{\delta_{(H^\pm,\, h^0)}}
\def\dhcw{\delta_{(H^\pm,\, W)}}
\def\dhw{\delta_{(h^0,\, W)}}
\def\dtb{\delta_{(t,\, b)}}
\def\dec{H^+\to W^+\gamma}
\begin{document}

\title{Decay $H^+\to W^+\gamma$ in a nonlinear $R_\xi$-gauge}
\author{J. Hern\'andez-S\'anchez}
\affiliation{Centro de Investigaci\'on Avanzada en Ingenier\'{\i}a
Industrial, Universidad Aut\'onoma del Estado de Hidalgo, C.P.
42184, Pachuca, Hidalgo, M\'exico}
\author{M. A. P\'erez}
\affiliation{Departamento de F\'isica, CINVESTAV, Apartado Postal
14-740, 07000, M\'exico, D. F., M\'exico}
\author{G. Tavares-Velasco}
\affiliation{Facultad de Ciencias F\'isico Matem\'aticas,
Benem\'erita Universidad Aut\'onoma de Puebla, Apartado Postal
1152, Puebla, Pue., M\'exico}
\author{J. J. Toscano}
\affiliation{Facultad de Ciencias F\'isico Matem\'aticas,
Benem\'erita Universidad Aut\'onoma de Puebla, Apartado Postal
1152, Puebla, Pue., M\'exico}

\date{\today}

\begin{abstract}
A new evaluation of the charged Higgs boson decay $H^+\to
W^+\gamma$ is presented in the context of the general two-Higgs
doublet model. A nonlinear $R_\xi$-gauge which considerably
simplifies the calculation is introduced and simple expressions
are obtained for the fermionic and bosonic contributions. The
$H^+\to W^+\gamma$ branching ratio is analyzed for several values
of the parameters of the model. Although this decay can have a
branching fraction as large as $10^{-4}$ in a certain region of
the parameter space, it is found that such a region is disfavored
by the most recent constraints on $b \to s\gamma$, $g-2$ of the
muon, $Z\to b \bar b$, and the $\rho$ parameter, along with the
exclusions from direct searches at the CERN $e^-e^+$ LEP collider.
The possibility of detecting this decay at future colliders is
discussed.
\end{abstract}


\maketitle

\section{Introduction}
\label{int}One of the most popular extensions of the standard
model (SM) is the two-Higgs doublet model (THDM), mainly because
of its simplicity and consistency with the minimal realization of
supersymmetry (SUSY) \cite{HHG}. The THDM predicts the existence
of five Higgs bosons: two neutral CP-even scalar bosons $h^0$ and
$H^0$, one neutral CP-odd scalar boson $A^0$, and a pair of
charged scalar bosons $H^\pm$ \cite{HHG}. The detection of the
charged Higgs boson would be an irrefutable signal of an extended
Higgs sector, and its experimental study might help us to
elucidate the structure of the Higgs potential and shed light on
the group representation of the Higgs fields through the study of
its couplings to the gauge boson fields. In models which only
comprise singlets and doublets of Higgs bosons, the decays of the
charged scalar boson into a gauge boson pair, namely $H^+\to
W^+\gamma$ \cite{Capdequi,Raychaudhury,Wudka} and $H^+\to W^+Z$
\cite{Mendez,Kanemura}, are absent at the tree level, though they
may occur at higher orders. While the decay mode $H^+\to
W^+\gamma$ is forbidden at the tree level due to electromagnetic
gauge invariance, the $H^+\to W^+Z$ decay can be induced at this
order in models including Higgs triplets or more complicated
representations \cite{HHG}. In spite of their suppressed branching
ratios, these decay modes are very interesting, which is due to
the fact that their experimental study may provide important
information concerning the underlying structure of the gauge and
scalar sectors. Apart from being rather sensitive to new physics
effects \cite{Diaz}, these channels have a clear signature and
might be at the reach of future particle colliders.

In this work we present an evaluation of the one-loop induced
decay $H^+\to W^+\gamma$ in the framework of the THDM. This decay
was already studied in the context of the minimal supersymmetric
standard model (MSSM), which is a restricted case of the general
THDM. At the one-loop, the main contributions to this decay arise
from the third-generation quarks $t$ and $b$, whereas the bosonic
sector contributes through the CP-even Higgs bosons $h^0$ and
$H^0$ along with the $W$ boson. The fermionic contribution and the
effects of a fourth family were studied in Ref. \cite{Capdequi}.
The bosonic contribution was studied in the linear $R_\xi$-gauge
\cite{Raychaudhury}, which is not suitable for such a calculation
since gives rise to several complications that can be evaded by
the use of a nonlinear $R_\xi$-gauge \cite{Fujikawa}. For
instance, while in the linear $R_\xi$-gauge one has to deal with
about 100 Feynman diagrams when computing the bosonic contribution
\cite{Raychaudhury}, in a nonlinear $R_\xi$-gauge one only needs
to compute 21 diagrams. As for the unitary gauge, the calculation
happens to be extremely hard due to the terms arising from the
longitudinal part of the gauge boson propagators \cite{Wudka}. Our
aim in this work is to reevaluate the $\dec$ decay in a nonlinear
$R_\xi$-gauge, which leads to considerably simplifications due to
the fact that some unphysical vertices are removed from the
interaction Lagrangian. We will see that such a gauge not only
reduces considerably the number of Feynman diagrams but also
renders manifestly gauge-invariant and ultraviolet-finite
amplitudes. Apart from emphasizing the advantages of using the
nonlinear $R_\xi$-gauge, we will analyze the $\dec$ decay in some
scenarios which are still consistent with the most recent bounds
obtained from electroweak precision measurements. Indeed, it has
been argued quite recently \cite{Cheung} that the parameter space
of the type-II THDM has become tightly constrained according to
the last reported value of the muon anomalous magnetic moment
$a_\mu$ \cite{muon}, so we will concentrate essentially on the
still-allowed region.

The rest of the paper is organized as follows. A brief review of
the model and the nonlinear $R_\xi$-gauge used in our calculation
are presented in Sec. \ref{model}. Sec. \ref{decwidth} is devoted
to discuss the most interesting details of the calculation with
emphasis on the facilities brought about by the nonlinear
$R_\xi$-gauge. The discussion and conclusions are presented in
Sec. \ref{discussion} and \ref{conclusions}, respectively.
Finally, the Feynman rules necessary for the calculation are shown
in Appendix \ref{appendix1}.

\section{The CP-conserving Higgs potential in the two-Higgs doublet model}
\label{model}The scalar sector of the CP-conserving THDM consists
of two scalar doublets with hypercharge $+1$:
$\Phi^\dag_1=(\phi^-_1,\phi_1^{0*})$ and
$\Phi^\dag_2=(\phi^-_2,\phi_2^{0*})$. The most general gauge
invariant potential can be written as \cite{Haber}
\begin{eqnarray}
V(\Phi_1,\Phi_2)&=&\mu^2_1(\Phi_1^\dag
\Phi_2)+\mu^2_2(\Phi^\dag_2\Phi_2)-\left(\mu^2_{12}(\Phi^\dag_1\Phi_2)+{\rm
H.c.}\right)+
\lambda_1(\Phi^\dag_1\Phi_1)^2+\lambda_2(\Phi^\dag_2\Phi_2)^2+\lambda_3(\Phi_1^\dag
\Phi_1)(\Phi^\dag_2\, \Phi_2)\nonumber \\
&+&\lambda_4(\Phi^\dag_1\Phi_2)(\Phi^\dag_2\Phi_1)+
\frac{1}{2}\left(\lambda_5(\Phi^\dag_1\Phi_2)^2+\left(\lambda_6(\Phi_1^\dag
\Phi_1)+\lambda_7(\Phi^\dag_2\Phi_2)\right)(\Phi_1^\dag \Phi_2)+
{\rm H.c}.\right) \label{potential}
\end{eqnarray}

\noindent It has been customary to impose the discrete symmetry
$\Phi_1\to \Phi_1$ and $\Phi_2\to -\Phi_2$ in order to avoid
dangerous flavor changing neutral current (FCNC) effects. This
symmetry is strongly violated by the terms associated with
$\lambda_6$, and $\lambda_7$, but it is softly violated by that
associated with $\mu^2_{12}$.  It is worth mentioning that the
last term plays an important role in SUSY models. All of the terms
in Eq. (\ref{potential}) are essential to obtain the decoupling
limit of the model, in which only one CP-even scalar boson is
light. In this work we will assume that this discrete symmetry is
softly violated, which means that we will take
$\lambda_6=\lambda_7=0$ throughout the rest of the presentation.

The scalar potential (\ref{potential}) has to be diagonalized to
yield the mass-eigenstates fields. The charged components of the
doublets lead to the physical charged Higgs boson and the
pseudo-Goldstone boson associated with the $W$ gauge field as
follows

\begin{eqnarray}
H^\pm&=&-\phi^\pm_1\sin\beta+\phi^\pm_2\cos\beta,\\
G^\pm_W&=&\phi^\pm_1\cos\beta+\phi^\pm_2\sin\beta,
\end{eqnarray}

\noindent with $\tan\beta=v_2/v_1$, being $v_1\;\, (v_2)$ the
vacuum expectation value (VEV) associated with $\Phi_1\,(\Phi_2)$.
On the other hand, the imaginary part of the neutral components
$\phi^0_{iI}$ define the neutral CP-odd scalar and the
pseudo-Goldstone boson associated with the $Z$ gauge boson. The
corresponding rotation is given by

\begin{eqnarray}
A^0&=&-\phi^0_{1I}\sin\beta+\phi^0_{2I}\cos\beta,\\
G_Z&=&\phi^0_{1I}\cos\beta+\phi^0_{2I}\sin\beta,
\end{eqnarray}
whereas the real part of the neutral components $\phi^0_{iR}$
define the CP-even Higgs bosons $h^0$ and $H^0$:

\begin{eqnarray}
h^0&=&-\phi^0_{1R}\sin\alpha+\phi^0_{2R}\cos\alpha,\\
H^0&=&\phi^0_{1R}\cos\alpha+\phi^0_{2R}\sin\alpha,
\end{eqnarray}

\noindent where the mixing angle $\alpha$ is given by

\begin{equation}
\tan 2\alpha=\frac{2\,m_{12}}{m_{11}-m_{22}},
\end{equation}

\noindent with the elements of the mass matrix being
$m_{11}=2v^2_1\lambda_1+\tan\beta\mu^2_{12}$,
$m_{22}=2v^2_2\lambda_2+\cot\beta\mu^2_{12}$, and
$m_{12}=v_1v_2(\lambda_3+\lambda_4+\lambda_5)-\mu^2_{12}$.

The masses of the charged and CP-odd scalar bosons satisfy the
following expression

\begin{equation}
m^2_{H^\pm}=m^2_{A^0}+\frac{2m^2_W}{g^2}(\lambda_5-\lambda_4),
\end{equation}

\noindent where $g$ is the coupling constant associated with the
$SU_L(2)$ group. Obviously $m_{H^\pm}=m_{A^0}$ when
$\lambda_5=\lambda_4$, which is a reflect of the underlying
custodial symmetry.

We now would like to comment on the gauge-fixing procedure which
was used to simplify our calculation. To this end we introduce the
following gauge-fixing functions \cite{Fujikawa}:

\begin{eqnarray}
f^+&=&\left(D^e_\mu +\frac{igs^2_W}{c_W}Z_\mu \right)W^{+\mu}-i\xi m_WG^+_W, \\
f^Z&=&\partial_\mu Z^\mu-\xi m_ZG_Z,\\
f^A&=&\partial_\mu A^\mu,
\end{eqnarray}

\noindent with $D^e_\mu$ the electromagnetic covariant derivative
and $\xi$ the gauge parameter. Note that $f^+$ is nonlinear and
transforms covariantly under the electromagnetic gauge group. This
gauge-fixing procedure is suited to remove the unphysical vertices
$WG_W\gamma$ and $WG_WZ$, which arise in the Higgs kinetic-energy
sector, and also modifies the Yang-Mills sector. One important
result is that the expression for the $WW\gamma$ vertex satisfies
a QED-like Ward identity, which turns out to be very useful in
loop calculations. In particular, in the Feynman-t'Hooft gauge the
Feynman rule for the $W^+_\rho(q)W^-_\nu (p) A_\mu (k)$ coupling
can be written as

\begin{equation}
\Gamma^{WW\gamma}_{\rho \nu \mu}=-ie\left(g_{\mu
\nu}(p-k+q)_\rho+g_{\nu \rho}(q-p)_\mu+g_{\mu
\rho}(k-p-q)_\nu\right),
\end{equation}
where all the momenta are incoming. It is easy to see that this
expression satisfies a QED-like Ward identity.

The remaining Feynman rules necessary for our calculation do not
depend on the gauge fixing procedure. For completeness they are
presented in Appendix \ref{appendix1}. As far as the Yukawa
couplings are concerned, we will concentrate on the type-II THDM,
in which the up-type quarks couple to the first Higgs doublet
whereas the remaining fermions couple to the second one.

\section{The $H^+\to W^+\gamma$ decay width}
\label{decwidth}

\begin{figure}[!hbt]
\centering
\includegraphics[width=3.in]{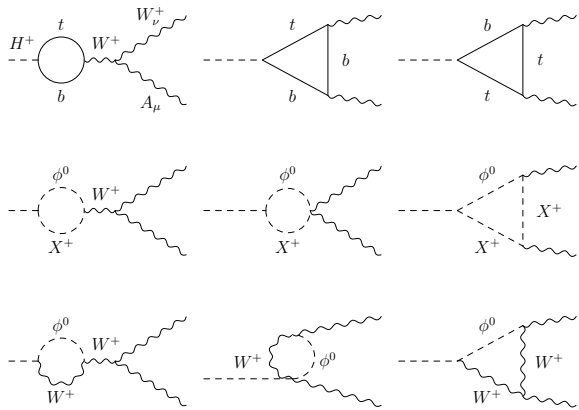}
\caption{\label{hwgdiag}Feynman diagrams contributing to the
$H^+\to W^{+}\gamma$ decay in the nonlinear $R_\xi$-gauge.
$\phi^0$ stands for $h^0$ and $H^0$, and $X^+$ for $H^+$ and
$G_W^+$. Compare with Fig. 1 of Ref. \cite{Raychaudhury}.}
\end{figure}

In the nonlinear $R_\xi$-gauge, the decay $H^+\to W^+\gamma$
receives contributions from the Feynman diagrams shown in Fig.
\ref{hwgdiag}. As far as the fermionic sector is concerned, the
main contribution comes from the third-generation quarks, which
induce three diagrams, whereas in the bosonic sector there are
contributions from the ($H^\pm,\,\phi^0$) and ($W^\pm,\,\phi^0$)
pairs, with $\phi^0=h^0$ or $H^0$. We would like to emphasize that
the nonlinear $R_\xi$-gauge considerably simplifies the
calculation of the $\dec$ decay. First of all, the removal of the
unphysical vertex $W^\pm G^\mp_W \gamma$, allows one to get rid of
all those diagrams which are displayed in Fig. 2 and do have to be
calculated in the linear $R_\xi$-gauge. Even more, in the
nonlinear $R_\xi$-gauge the tadpole graphs shown in Fig.
\ref{tadpole} vanish. Apart from these simplifications, we will be
able to group the Feynman diagrams into subsets which separately
yield a manifestly gauge-invariant amplitude free of ultraviolet
singularities.

\begin{figure}[!hbt]
\centering
\includegraphics[width=2.in]{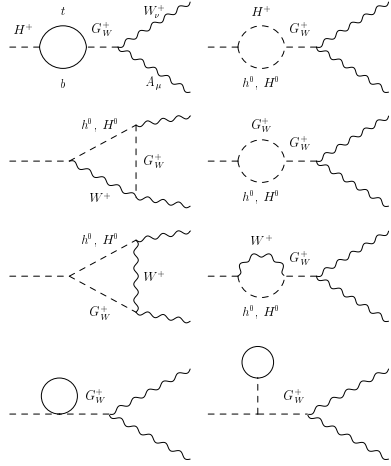}
\caption{\label{lghwgdiag}Feynman diagrams  which appear in the
linear $R_\xi$-gauge but are absent in the nonlinear one. $h_0$,
$H_0$, $A_0$, $H^\pm$, $G_W^\pm$ and $G_Z$ circulate in both
tadpoles, whereas $W^\pm$, $Z$, $t$ and $b$ also circulate in the
right-hand tadpole.}
\end{figure}

\begin{figure}[!hbt]
\centering
\includegraphics[width=2.5in]{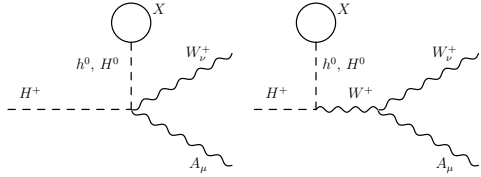}
\caption{\label{tadpole}Tadpole diagrams whose amplitude vanishes
in the nonlinear $R_\xi$-gauge. $X$ stands for $h^0$, $H^0$,
$A^0$, $H^\pm$, $G_W^\pm$, $G_Z$, $W^\pm$, $Z$, $t$, and $b$.}
\end{figure}

Once the amplitude for every Feynman diagram shown in Fig.
\ref{hwgdiag} was written down, the Passarino-Veltman method
\cite{Passarino} was applied to express it in terms of scalar
integrals, which are suitable for numerical evaluation \cite{FF}.
The full amplitude for the decay $H^+(p)\to W^+(q) \gamma (k)$ can
be written as

\begin{equation}
\mathcal{M} ( H^+ \to W^+ \gamma  ) =
 \frac{\alpha^{3/2}}{2 \sqrt{\pi} m_W s_w^2}
 \left( G\,\left(q\cdot k \;g_{\mu\nu}-
  k_{\nu} q_\mu\right) - i\, H\, \epsilon_{\mu \nu \alpha \beta}
k^{\alpha}  q^\beta \right)\epsilon^{\mu } (k) \epsilon^{\nu }
(q), \label{amplitude}
\end{equation}

\noindent which is manifestly gauge invariant. The $H$ function
only receives contributions from the quarks:

\begin{equation}
H = \frac{i\,\lambda_{t b}}{\dhcw} \left(3\, \delta
B_{(H^\pm,W,b,t)}+\dhcw\left(c_\beta^2\, C_{(b,t)}+2\,s_\beta^2\,
C_{(t,b)}\right)\right),
\end{equation}

\begin{equation}\lambda_{tb}=-\frac{m_b\,m_t}{s_\beta
c_\beta},\end{equation}

\noindent where we have introduced the shorthand notation
$\delta_{(a,\,b)}\equiv m_a^2-m_b^2$, $\delta
B_{(a,b,c,d)}=B_0\left(m_a^2,m_c^2,m_d^2\right)-
B_0\left(m_b^2,m_c^2,m_d^2\right)$,\footnote{When $a=0$, $m_a$
must be substituted by 0.} and $C_{(a,b)}\equiv
C_0\left(m^2_{H^\pm},m_W^2,0,m_a^2,m_b^2,m_a^2\right)$, where
$B_0$ and $C_0$ stand for Passarino-Veltman scalar integrals. In
addition, $c_\beta=\cos\beta$, $s_\beta=\sin\beta$, etc. From the
$\delta B_{(a,b,c,d)}$ definition, it is clear that the $H$
function is ultraviolet finite. As for $G$, it can be written as

\begin{eqnarray}
G=\frac{1}{2\,\mhc^2\, \dhcw^2}\left(G_{tb} +\sum_{\phi^0=H^0,\,
h^0}\left(G_{ H^\pm\phi^0}+ G_{G_W^\pm \phi^0}+ G_{W^\pm
\phi^0}\right)\right), \label{Gdef}
\end{eqnarray}

\noindent where $G_{AB}$ stands for the contribution of the
$(A,\,B)$ pair:

\begin{eqnarray}
G_{tb} & = &\lambda_{tb}
\Big(\mhc^2\left(3\,\left(\dtb+c_{2\beta}\,\dhcw\right)+m_W^2\right)\delta
B_{(H^\pm,W,b,t)}-3\,\dtb\dhcw \delta
B_{(0,H^\pm,b,t)}\nonumber\\&+&\mhc^2\dhcw
\left(1-\left(2\,m_b^2-c_\beta^2\dhcw\right)C_{(b,t)}+\left(2\,m_t^2-s_\beta^2\,\dhcw\right)
C_{(t,b)}\right)\Big),
\end{eqnarray}

\begin{eqnarray}
G_{H^\pm h^0 } &=&\lambda_{{h^0 H^\pm}}\Big(\dhch\dhcw \delta
B_{(0,W,H^\pm,h^0)}- \left(2\,\mhc^2\dhch+m_{h^0}^2
m_W^2\right)\delta
B_{(H^\pm,W,H^\pm,h^0)}\nonumber\\&-&\mhc^2\dhcw\left(1+2\,\mhc^2
C_{(H^\pm,h^0)}\right)\Big),
\end{eqnarray}

\begin{eqnarray}
G_{G_W^\pm h^0} &=&\lambda_{{G_W^\pm h^0}}\Big(
\left(m_W^2\dhw+\mhc^2(3m_W^2-2m_{h^0}^2)\right)\delta
B_{(H^\pm,W,h^0,W)}+\dhch\,\dhw \delta
B_{(0,W,h^0,W)}\nonumber\\&+& \mhc^2\dhcw\left(1+2\,m_W^2
C_{(W,h^0)}\right) \Big),
\end{eqnarray}
and
\begin{eqnarray}
G_{W^\pm H^0} & = &\lambda_{W^\pm h^0} \Big(
\left(m_W^2\dhw+\mhc^2\left(7\,m_W^2-2\,m_{h^0}^2-4\,\mhc^2\right)\right)\delta
B_{(H^\pm,W,H^\pm,h^0)}+\dhcw\dhw \delta
B_{(0,W,H^\pm,h^0)}\nonumber\\&+&\mhc^2\dhcw\left(1-2\,\left(2\,\dhcw-m_W^2\right)C_{(m_W^2,m_{h^0}^2)}\right)
\Big),
\end{eqnarray}
with

\begin{equation}\lambda_{{ H^\pm h^0}}=c_{\alpha-\beta}\left(2\,s_{ \alpha - \beta }\,m_{H^\pm}^2
 +\frac{ c_{\alpha + \beta}}{c^2_{\beta }s^2_\beta}\,\mu_{12}^2
 + \,\left(s_\alpha s_\beta t_\beta - \frac{c_\beta
c_\alpha}{t_\beta}\right)\,m_{h^0}^2 \right),\end{equation}

\begin{equation}\lambda_{{G_W^\pm h^0}}=\frac{s_{2( \alpha -\beta)}\dhch}{2},\end{equation}
and
\begin{equation}\lambda_{W^\pm h^0}= -\frac{m_W^2 s_{2( \alpha
-\beta)}}{2}.\end{equation}

\noindent Finally, the contribution of the heaviest CP-even scalar
boson $H^0$ is obtained from that of the lightest one once the
substitutions $m_{h^0}\to m_{H^0}$ and $\alpha\to\alpha-\pi/2$ are
done.

The above expressions are very simple and should be compared to
those presented in Refs. \cite{Capdequi,Raychaudhury}. It is also
evident that the partial amplitudes induced by the pairs
$(t,\,b)$, $(\phi^0,\, W^\pm)$, $(\phi^0,\, H^\pm)$, and
$(\phi^0,\, G_W^\pm)$ are gauge invariant and ultraviolet finite
on their own. Again, this is to be contrasted with the situation
arising in the linear $R_\xi$-gauge, where showing gauge
invariance is somewhat cumbersome, and the cancellation of
ultraviolet divergences in the bosonic sector is achieved only
after adding up all the Feynman diagrams \cite{Raychaudhury}.

Once Eq. (\ref{amplitude}) is squared and the spins of the final
particles are summed over, the decay width can be written as

\begin{equation}
\Gamma(\dec) =  \frac{\alpha^3\,\dhcw^3}{2^7 \pi^2 s_W^4 m^2_W
\mhc^3}  \left( |G|^2+|H|^2\right).
\end{equation}

We will evaluate this decay width for some values of the
parameters of the model.

\section{Numerical results and discussion}
\label{discussion}

In order to evaluate the $\dec$ decay, there are six free
parameters of the THDM which should be addressed. They are the
masses of the four scalar bosons $m_{H^\pm}$, $m_{h^0}$, $m_{H^0}$
and $m_{A^0}$, together with the VEVs ratio $\tan\beta$ and the
mixing angle $\alpha$. Without loosing generality, we will assume
that $\mu_{12}=0$. Although the $\dec$ decay width does not depend
on $m_{A^0}$, the respective branching ratio does depend on all of
these parameters. Instead of examining the behavior of $\dec$ for
arbitrary values of these parameters and looking for those which
enhance its branching ratio, a realistic analysis must take into
account the most recent constraints obtained from precision
measurements and direct searches at particle colliders. We will
thus concentrate on the region of the parameter space which is
still consistent with low-energy data.

The parameter space of THDMs as constrained by electroweak
precision measurements has been recurrently studied in the
literature \cite{Bounds,Cheung,Larios}. Along this line, it is
well known that the experimental bound on the inclusive decay
$b\to s\gamma$ can be translated into a bounded area on the
$m_{H^\pm}$-$\tan\beta$ plane. In addition, the muon anomalous
magnetic moment $a_\mu$ is useful to constrain the CP-odd scalar
boson mass $m_{A^0}$, which along with the CP-even scalar boson
masses $m_{h^0}$ and $m_{H^0}$ can be further constrained by the
$Z\to b\bar{b}$ decay mode $R_b$. All these constraints, when
combined with those obtained from the $\rho$ parameter and the
data from direct searches at the CERN $e^-e^+$ LEP collider, can
yield the allowed region in the parameter space of THDMs. Quite
recently, the most up-to-date measurements on $b\to s \gamma$
\cite{btosgamma}, $a_\mu$ \cite{muon}, $R_b$ \cite{Rb}, and $\rho$
\cite{PDG}, together with the DELPHI \cite{DELPHI} and OPAL
\cite{OPAL} excluded regions, were considered to constrain the
parameter space of the type-II THDM \cite{Cheung,Larios}. In
particular, the possibility of a very light CP-odd scalar boson
was considered in \cite{Larios}, whereas a more general treatment
was presented in \cite{Cheung}. From these studies one is lead to
conclude that the parameter space of the THDM has become tightly
constrained by the most recent measurement of $a_\mu$ \cite{muon}.
Below we will follow closely the analysis presented in Ref.
\cite{Cheung}.

The model independent bound $m_{H^\pm}\gtrsim 80$ GeV has been
derived at LEP for a charged Higgs boson decaying solely into
$\tau\bar{\nu}$ \cite{L3Bound}. On the other hand, within the
THDM-II, the latest reported value for the decay $b\to s \gamma$
\cite{btosgamma} yields the very stringent lower bound
$m_{H^\pm}\ge 500$ GeV for intermediate and large values of $\tan
\beta$, though this bound can loosen in SUSY models with conserved
or broken $R$-parity \cite{Goto}. As far as $a_\mu$ is concerned,
the latest experimental data from the BNL collaboration
\cite{muon} require a positive contribution from new physics to
bring the theoretical value closer to the experimental one. It
turns out that the CP-odd scalar yields a positive contribution to
$a_\mu$, whereas the remaining Higgs bosons give a negative
contribution. This means that the deviation between the
experimental and theoretical values of $a_\mu$ can only be
explained by the existence of a light CP-odd scalar along with a
high value of $\tan\beta$ and relatively heavy $H^\pm$, $h^0$ and
$H^0$. When the $R_b$ constraint \cite{Rb} is taken into account,
a large region of the parameter space of the THDM gets excluded,
and only a tiny region survives when the DELPHI \cite{DELPHI} and
OPAL \cite{OPAL} excluded regions for the scalar boson masses are
considered. From this analysis, it was concluded \cite{Cheung}
that the most favorable region is that in which $\tan \beta$ is of
the order of 50 or larger, $m_{A^0}\le 80$ GeV, $m_{h^0}\le 140$
GeV, and $ -1/2\,\pi\le \alpha \le -3/8\,\pi$. Even more, the
$\rho\sim 1$ constraint requires fine-tuned solutions for the
masses of the scalar bosons and the mixing angles. In fact,  this
constraint is satisfied by an almost decoupling CP-even scalar
boson $H^0$, with a mass of the order of 1 TeV, provided that the
remaining parameters meet the previous constraints.

Bearing in mind the preceding discussion, we will analyze the
$\dec$ decay for 300 GeV $\le m_{H^\pm}\le$ 600 GeV. For the sake
of illustration we will assume that the lightest CP-even scalar
boson is relatively light and the heaviest one is much heavier
indeed, {\it i.e.} we will take $m_{h^0}=115$ GeV and $m_{H^0}=1$
TeV. The former value is consistent with the current LEP bound on
the mass of a SM-like Higgs boson \cite{PDG}. Moreover, we will
focus on the region where $\tan\beta$ is large and $\sin\alpha$ is
close to -1. We have verified that these values lie inside the
still-allowed region of the parameter space of the THDM-II.

\begin{figure}[!hbt]
\includegraphics[width=3.5in]{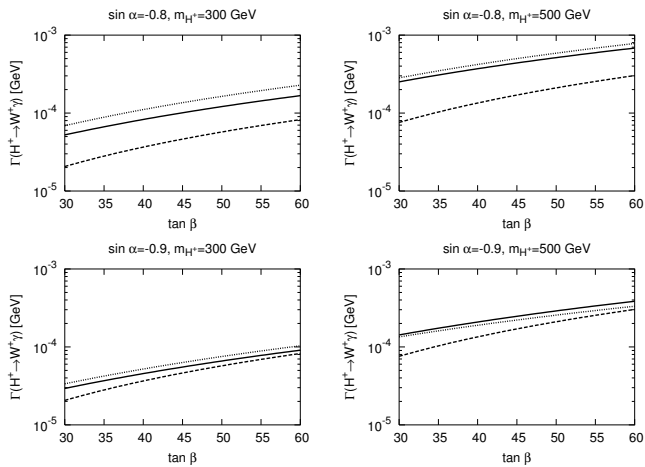}
\caption{\label{dectb} $H^+\to W^+\gamma$ decay width (solid line)
as a function of $\tan \beta$ for the indicated values of $\sin
\alpha$ and $m_{H^\pm}$. We have taken $m_{h^0}=115$ GeV and
$m_{H^0}=1$ TeV. The partial contributions from the quark (dashed
line) and boson (dotted line) sectors are also shown.}
\end{figure}

\begin{figure}[!hbt]
\centering
\includegraphics[width=3.5in]{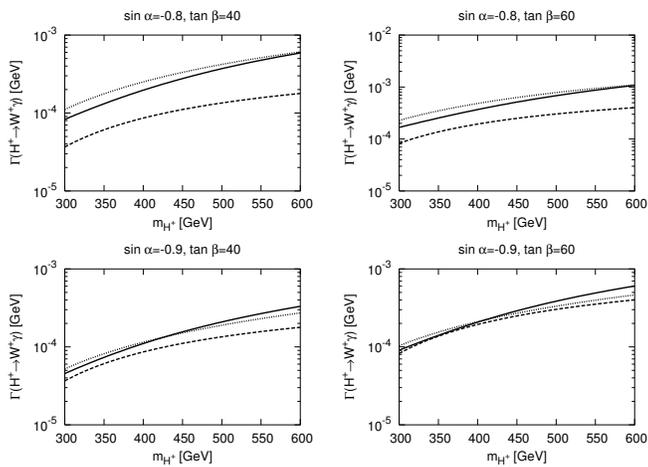}
\caption{\label{decmh} The same as Fig. \ref{dectb} but now as a
function of $m_{H^\pm}$.}
\end{figure}

\begin{figure}[!hbt]
\includegraphics[width=3.in]{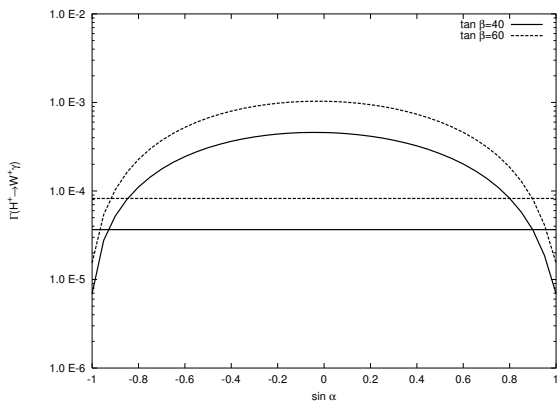}
\caption{\label{decsa} Partial  contributions to the $H^+\to
W^+\gamma$ decay width as a function of $\sin \alpha$ for
$m_{H^\pm}=300$ GeV and two different values of $\tan\beta$. The
constant lines correspond to the fermion contribution and the
remaining to the bosonic one. The values of the CP-even scalar
boson masses are the same as in Fig. \ref{dectb}.}
\end{figure}

It is interesting to examine the effects of both the quark and
boson sectors on the decay width $\Gamma(\dec)$. The partial
contributions, together with the full contribution, are displayed
as a function of $\tan\beta$ in Fig. \ref{dectb}, whereas  in Fig.
\ref{decmh} we show them as a function of $m_{H^\pm}$. We can
observe that, in this range of parameters, the $\dec$ decay is
slightly dominated by the bosonic contribution. The fermionic
contribution is larger only in the region of the parameter space
where $\sin\alpha$ is very close to -1. This is illustrated in
Fig. \ref{decsa}, where we show the partial contributions to the
$\dec$ decay as a function of $\sin\alpha$ for $m_{H^\pm}=$ 300
GeV and two different values of $\tan\beta$. We see that the
bosonic contribution is maximal when $\sin\alpha=0$ and minimal
when $|\sin\alpha|=1$, where it is smaller than the fermionic
contribution. For this particular set of the parameters, the
maximal value of the bosonic contribution is more than one order
of magnitude than the respective minimal value. A similar behavior
is observed for larger values of $m_{H^\pm}$ and $\tan\beta$.
Also, from Fig. \ref{dectb} and Fig. \ref{decmh} it is interesting
to note that the partial contribution from the bosonic sector
exceeds the total contribution in some regions of the parameter
space. This stems from the fact that the interference between the
partial contributions is destructive. While the fermion
contribution depends only on $\tan\beta$ for a given $m_{H^\pm}$,
the bosonic contribution depends on four other unknown parameters.
This makes harder the analysis of this decay mode.

\begin{figure}[!hbt]
\centering
\includegraphics[width=3.in]{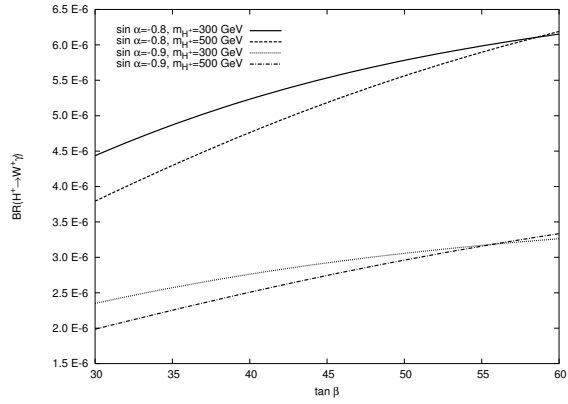}
\caption{\label{brtb} Branching ratio for the $H^+\to W^+\gamma$
decay as a function of $\tan\beta$ and the indicated values of
$\sin\alpha$ and $m_{H^\pm}$.}
\end{figure}

\begin{figure}[!hbt]
\centering
\includegraphics[width=3.in]{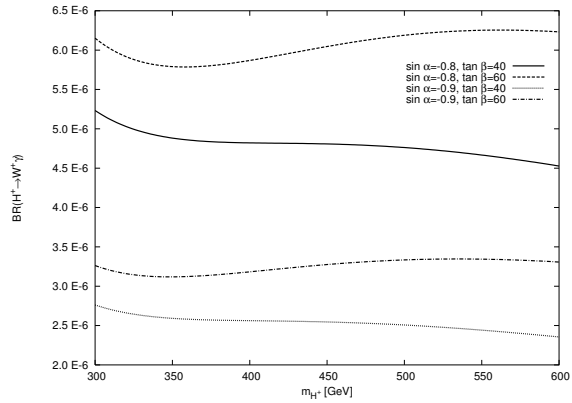}
\caption{\label{brmh} Branching ratio for the $H^+\to W^+\gamma$
decay as a function of $m_{H^\pm}$ for the indicated values of
$\sin\alpha$ and $\tan\beta$.}
\end{figure}

The full width of the charged scalar boson is given to a good
approximation by the sum of the partial widths of the two-body
decay channels induced at the tree-level: $H^+\to \tau \bar \nu$,
$c\bar s$, $t \bar b$, $W^+ A^0$, and $W^+h^0$.  The widths for
all these decays can be readily obtained and we refrain from
presenting the respective expressions here. In the region of the
parameter space that we are considering, the charged scalar boson
would decay mainly into $t\bar{b}$. This decay mode is somewhat
troublesome as it would suffer from large QCD backgrounds at a
hadronic collider \cite{Barger}. The $H^+\to W^+\gamma$ branching
ratio is shown in Fig. \ref{brtb} as a function of $\tan\beta$ for
$m_{A^0}=80$ GeV and two different values of $\sin\alpha$. We can
observe that $Br(\dec)$ is at most of the order of
$6\times10^{-6}$, it increases for larger values of $\tan\beta$,
but falls slowly as $\sin\alpha$ gets closer to $-1$. On the other
hand, the behavior of $Br(\dec)$ as a function of $m_{H^\pm}$ is
depicted in Fig. \ref{brmh}. For $\tan\beta=40$, the branching
fraction decreases very slowly for larger $m_{H^\pm}$, whereas for
$\tan\beta=60$ it has a minimal around $m_{H^\pm}=350$ GeV and
increases slowly until about $m_{H^\pm}=550$ GeV, where there is a
maxima.

\begin{figure}[!hbt]
\centering
\includegraphics[width=3.in]{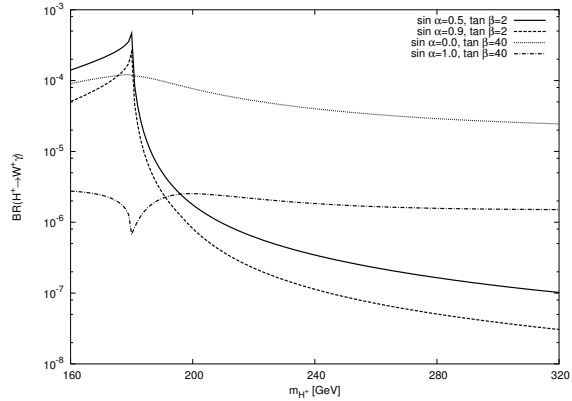}
\caption{\label{brmh-2} The same as Fig. \ref{brmh} in other
region of the parameter space of the THDM-II.}
\end{figure}

It is worthwhile to mention that the $\dec$ branching ratio may be
dramatically enhanced, mainly by the fermionic contribution,
around the threshold $m_{H^\pm}\sim m_t+m_b$. This is shown in
Fig. \ref{brmh-2} for some illustrative values of $\tan\beta$ and
$\sin\alpha$. It can be observed that $Br(\dec)$ can be larger
than $10^{-4}$ when $\tan\beta$ is small regardless of the value
of $\sin\alpha$, and up to $10^{-4}$ when a large value of
$\tan\beta$ combines with $\sin\alpha\sim 0$. In the latter case
the bosonic contribution is also responsible for the enhancement.
However, when a large value of $\tan\beta$ combines with
$\sin\alpha\sim 1$, there is large cancellations between the two
contributions, which results in a significant suppression of
$Br(\dec)$. Unfortunately the region of the parameter space in
which such an enhancement shows up is not favored by the
constraints from electroweak precision measurements. Although a
fourth-fermion family may improve the $\dec$ decay rate
\cite{Capdequi}, the enhancement would depend on the splitting
between the masses of the fourth-generation quarks
$m_{u^\prime}^2-m_{d^\prime}^2$, which cannot be too large due to
the $\rho\sim 1$ restriction. In this respect, it seems that the
$H^\pm\to W^\pm Z$ decay is more promising since it is sensitive
to nondecoupling effects when there is a large splitting between
the CP-odd and the charged scalar boson masses \cite{Kanemura}. As
discussed above, this scenario is still consistent with the
constraints on the parameter space of the THDM-II. Furthermore,
the $H^\pm\to W^\pm Z$ decay can arise at the tree-level in Higgs
triplet models \cite{HHG}.

Let us now discuss shortly on the possibility of detecting the
$\dec$ decay at future colliders. Single and pair production of
heavy charged Higgs bosons at both hadronic \cite{Barger,
H+H-,gbtotH,Various,cstoH+,taupol,H+W-} and linear $e^-e^+$
colliders \cite{e-e+,egamma,gamma-gamma} have long received
considerable attention. A charged scalar boson heavier than $180$
GeV could be detected at a hadronic collider via several
processes, such as $gb\to H^-t$ \cite{Barger,gbtotH}, $c\bar{s}\to
H^+$ \cite{cstoH+}, and $c\bar{b}\to H^+$ \cite{cstoH+}. The
one-loop processes $gg\to W^\pm H^\mp$ and $q\bar{q}\to W^\pm
H^\mp$ give no viable signature at the CERN LHC collider
\cite{H+W-}. While the $gb\to H^-t$ rate is out of the reach of
the Fermilab Tevatron, it could be detected at the LHC for
$m_{H^\pm}$ up to 1 TeV \cite{Barger,gbtotH}. This process is
appropriate to look for a charged Higgs boson in the low- and
high-$\tan\beta$ regime. Though the $gb\to H^-t$ process followed
by the decay $H^-\to t\bar{b}$ suffers from irreducible
background, the $H^-\to \tau \bar{\nu}$ mode is background free
due to the distinctive $\tau$ polarization \cite{taupol}. The
cross section for $gb\to H^-t$ increases for large $\tan\beta$ but
falls when $m_{H^\pm}$ increases. At the LHC, the cross section
for $pp\to H^+ \bar{t} b$ can be as large as $10^3$ fb for
$m_{H^\pm}$ of the order of a few hundreds of GeV and $\tan\beta$
around 50. With an integrated luminosity of 100 fb, this rate is
enough to detect a charged scalar boson decaying as $H^+\to
\bar{\tau}\nu$, but it is not viable to detect the $H^+\to
W^+\gamma$ mode unless its decay width is dramatically enhanced.
There is some chance to detect the $\dec$ decay if the constraints
on the parameter space can be evaded in some way and the charged
scalar boson mass is about 180 GeV. In this case the branching
ratio might increase up to $10^{-4}$, whereas the cross section
for $pp\to H^+ \bar{t} b$ can reach the $10^4$ fb level.

In a linear collider (LC) the situation is less promising. If the
mass of the charged scalar boson happen to be smaller than half
the center-of-mass energy $\sqrt{s}$ of the LC, it might be
detected through the double pair production process $e^-e^+\to
H^-H^+$ \cite{e-e+}, which has the advantage that it does not
depend on $\tan\beta$ as it is dictated entirely by the
interactions of the charged Higgs boson with the neutral gauge
bosons. On the other hand, when $\sqrt{s} < m_{H^\pm}/2$, a
charged scalar boson might be detected via the single production
modes $e^-e^+ \to \tau\bar{\nu}H^+$, $e^-e^+ \to b\bar{c}H^+$, or
$e^-e^+\to W^\pm H^\mp$ \cite{e-e+}. The first two are more
promising as arise at the tree level, whereas the last one occurs
at one loop. In a linear collider running at $\sqrt{s}=500$ GeV,
the cross section for $e^-e^+ \to \tau\bar{\nu}H^+$ would be about
$10^{-2}$ fb for $m_{H^\pm}=300$ and $\tan\beta=60$ GeV. Even with
an integrated luminosity of $500$ ${\rm fb}^{-1}$, this rate is
too small to allow the detection of the $\dec$ decay. If
$\sqrt{s}=1000$ GeV, $\sigma(e^-e^+ \to H^-H^+)\sim 10$ fb for
$m_{H^\pm}=300$ GeV, which evidently is also too small for the
$\dec$ mode to be detected.

Since a LC would also work in the mode $\gamma\gamma$, it has been
recently proposed to look for a heavy charged scalar boson, with
mass larger than half the center-of-mass energy of the LC, through
the single production processes $\gamma \gamma \to\tau \bar{\nu}
H^+$ and $\gamma\gamma\to b\bar{c}H^+$ \cite{gamma-gamma}.  The
viability of these modes depends strongly on the strength of the
Yukawa couplings of the fermions to the charged scalar boson. It
turns out that the rates for these processes can be up to two
orders of magnitude larger than for the respective $e^-e^+$
processes \cite{gamma-gamma}. At a $\sqrt {s}=500$ GeV linear
collider, the cross section for $\sigma(\gamma \gamma \to\tau
\bar{\nu} H^+)$ is of the order of 1 fb, which is a typical value
for intermediate $m_{H^\pm}$ and large $\tan\beta$. The detection
of $\dec$ would thus not be viable. Of course the charged Higgs
boson can also be pair produced in the $\gamma\gamma$ mode
provided that the respective center-of-mass energy is enough, but
the conclusion remains the same. A linear collider would thus not
be appropriate to look for rare decays of the charged Higgs boson.

We have mentioned that while the $\dec$ branching ratio may reach
the level of $10^{-4}$ provided that the mass of the charged
scalar boson is around the threshold $m_t+m_b$,  it is at most of
the order of $10^{-6}$ in the region of the parameter space
favored by electroweak precision measurements. Even if the $\dec$
decay rate was not reduced considerably by the kinematical cuts,
it would be necessary the production of about $10^6$ charged
scalar bosons to observe a few events. This seems to be beyond the
reach of the LHC, so the $\dec$ decay would hardly be detected
unless its decay width is dramatically enhanced in other model.

\section{Summary}
\label{conclusions}A study of the decay $H^+\to W^+\gamma$ was
presented in the context of the type-II THDM. A nonlinear
$R_\xi$-gauge covariant under the electromagnetic gauge group was
used to obtain the one-loop amplitudes. The advantages of using
such a gauge were emphasized. In particular, there are 21 Feynman
diagrams in the nonlinear $R_\xi$-gauge, whereas the linear one
induces about one hundred. An analysis of the behavior of the
$H^+\to W^+\gamma$ decay width as a function of the parameters of
the model was presented, but we focused essentially on the region
of the parameter space still allowed by the constraints obtained
from the latest reported values of $b\to s\gamma$, $R_b$, $a_\mu$
and the $\rho$ parameter, which favor a charged scalar boson
heavier than the $t$ quark, a large value of $\tan\beta$, and $-
\pi/2\le \alpha \le -3\pi/8$. In this still-allowed region, which
was not considered in the previous analyses of the $\dec$ decay,
the respective branching ratio receives contribution of the same
order of magnitude from both the bosonic and fermionic sectors,
and it can be as large as $10^{-6}$, which seems to be beyond the
reach of the LHC.

{\it Note added:} After the submission of this manuscript, the
muon $g-2$ collaboration announced a new result for the anomalous
magnetic moment of the negative muon, which is based on the data
collected during the year 2001 \cite{Bennett:2004pv}. The new
result, $a_{\mu^-}^{\rm{exp}}=11 659 214(8)(3) \times
10^{-10}\,(0.7\; {\mathrm{ppm}})$, is consistent with previous
measurements of the anomaly for the negative and positive muon.
The new world average for the muon anomaly is thus
$a_\mu^{\rm{exp}}=11 659 208(6) \times 10^{-10}\,(0.5\;
{\mathrm{ppm}})$, which is to be contrasted with the previous one
\cite{muon}: $a_\mu^{\rm{exp}}=11 659 203(8) \times
10^{-10}\,(0.7\; {\mathrm{ppm}})$. This new results confirms the
fact that THDMs are very disfavored by the muon anomaly. However,
we would like to point out that the scenario that we are
considering is still consistent with the new result. So, the
analysis presented in this work is not affected significantly and
our conclusions remain unchanged.

\acknowledgments{This work is supported by Conacyt and SNI
(M\'exico). G.T.V. also acknowledges partial support from
SEP-PROMEP and thanks C.-P. Yuan for useful comments.}

\appendix
\section{Feynman rules for the decay $H^+\to W^+
\gamma$} \label{appendix1} In this appendix we present the Feynman
rules which do not depend on the gauge fixing procedure described
in Sec. \ref{model}. The $H^\pm H^\mp h^0$ vertex arise from the
Higgs potential. In terms of the coupling constants $\lambda_i$
the Feynman rule for this vertex is given by
\begin{equation}
g_{H^\pm H^\mp h^0}=g_{A^0A^0h^0}-\frac{2 i
m_W}{g}(\lambda_5-\lambda_4)s_{\beta-\alpha},
\end{equation}

\noindent where

\begin{equation}
g_{A^0A^0h^0}=\frac{2im_W}{g}\left(2\,\lambda_1 s^2_\beta c_\beta
s_\alpha-2\lambda_2 c^2_\beta s_\beta c_\alpha
-\left(\lambda_3+\lambda_4+\lambda_5\right)(s^3_\beta
c_\alpha-c^3_\beta s_\alpha)+2\lambda_5 s_{\beta-\alpha}\right).
\end{equation}

The coupling constants $\lambda_i$ can be expressed in turn in
terms of the mixing angles and the Higgs boson masses as follows:

\begin{eqnarray}
\lambda_1&=&\frac{g^2}{8m^2_W}\left(\left(\frac{s_\alpha}{c_\beta}\right)^2m^2_{h^0}+
\left(\frac{c_\alpha}{c_\beta}\right)^2m^2_{H^0}-\frac{s_\beta}{c^3_\beta}\mu^2_{12}\right), \\
\lambda_2&=&\frac{g^2}{8m^2_W}\left(\left(\frac{c_\alpha}{s_\beta}\right)^2m^2_{h^0}+
\left(\frac{s_\alpha}{s_\beta}\right)^2m^2_{H^0}-\frac{c_\beta}{s^3_\beta}\mu^2_{12}\right), \\
\lambda_3&=&\frac{g^2}{4m^2_W}\left(2m^2_{H^\pm}+\frac{s_{2\alpha}}{s_{2\beta}}
(m^2_{H^0}-m^2_{h^0})-\frac{\mu^2_{12}}{s_\beta c_\beta}\right),\\
\lambda_4&=&\frac{g^2}{4m^2_W}\left(m^2_{A^0}-2\,m^2_{H^\pm}+\frac{\mu^2_{12}}{s_\beta c_\beta}\right), \\
\lambda_5&=&\frac{g^2}{4m^2_W}\left(\frac{\mu^2_{12}}{s_\beta
c_\beta}-m^2_{A^0}\right).
\end{eqnarray}

\noindent After some algebra we are left with

\begin{equation}
g_{H^\pm H^\mp h^0}=\frac{i\,g}{2\,m_W}\left(2\,s_{ \alpha - \beta
}\,m_{H^\pm}^2
 +\frac{c_{\alpha + \beta}}{c^2_{\beta }s^2_\beta}\,\mu_{12}^2
 + \,\left(s_\alpha s_\beta t_\beta - \frac{c_\beta
c_\alpha}{t_\beta}\right)\,m_{h^0}^2 \right).
\end{equation}

The Higgs kinetic-energy term induces the $H^\pm G^\mp_W\,h^0$
vertex:
\begin{equation}
g_{H^\pm
G^\mp_Wh^0}=\frac{ig}{2m_W}(m^2_{H^\pm}-m^2_{h^0})c_{\beta-\alpha}.
\end{equation}

The Feynman rules involving the heavy CP-even scalar boson $H^0$
can be obtained from above after the replacements $m_{h^0}\to
m_{H^0}$ and $\alpha\to \alpha-\pi/2$. As for the couplings of the
fermions to the $W$ gauge boson and the charged scalar boson, they
are \cite{HHG}:

\begin{eqnarray}
g_{H^- \,t\,\bar{b}}&=&\frac{i\,g}{2\,\sqrt{2}\,m_W}\left(v+a\,\gamma^5\right),\\
g_{W_\mu^-
\,t\,\bar{b}}&=&\frac{i\,g}{2\,\sqrt{2}\,m_W}\left(v+a\,\gamma^5\right)\gamma_\mu,
\end{eqnarray}
with $v=m_t\,\cot\beta+m_b\,\tan\beta$ and
$a=m_b\,\tan\beta-m_t\,\cot\beta$.

\end{document}